\documentclass{article}
\usepackage{spconf,amsmath,amsfonts,graphicx}
\usepackage{subcaption}
\usepackage{cite}
\usepackage{booktabs}
\usepackage{multirow}
\usepackage{array}
\usepackage[dvipsnames]{xcolor}
\usepackage{flushend}

\title{FILTERBANK LEARNING FOR NOISE-ROBUST SMALL-FOOTPRINT KEYWORD SPOTTING}
\name{Iv\'an L\'opez-Espejo$^{1,2}$, Ram C. M. C. Shekar$^2$, Zheng-Hua Tan$^1$, Jesper Jensen$^{1,3}$, John H. L. Hansen$^2$\thanks{This work has been funded by the European Union's Horizon 2021 research and innovation program under the Marie Skłodowska-Curie grant agreement No. 101062614.}}
\address{$^1$Department of Electronic Systems, Aalborg University, Denmark\\
		 $^2$Center for Robust Speech Systems (CRSS), The University of Texas at Dallas, USA\\
		 $^3$Oticon A/S, Denmark\\
		 \texttt{\small\{ivl,zt,jje\}@es.aau.dk, \{ramcharan.chandrashekar,john.hansen\}@utdallas.edu, jesj@demant.com}
}
\begin{document}
%
\maketitle
\begin{abstract}
In the context of keyword spotting (KWS), the replacement of handcrafted speech features by learnable features has not yielded superior KWS performance. In this study, we demonstrate that filterbank learning outperforms handcrafted speech features for KWS whenever the number of filterbank channels is severely decreased. Reducing the number of channels might yield certain KWS performance drop, but also a substantial energy consumption reduction, which is key when deploying common always-on KWS on low-resource devices. Experimental results on a noisy version of the Google Speech Commands Dataset show that filterbank learning adapts to noise characteristics to provide a higher degree of robustness to noise, especially when dropout is integrated. Thus, switching from typically used 40-channel log-Mel features to 8-channel learned features leads to a relative KWS accuracy loss of only 3.5\% while simultaneously achieving a 6.3$\times$ energy consumption reduction.
\end{abstract}
\section{Introduction}
\label{sec:intro}

Over recent years, there has been an increasing interest in developing end-to-end deep learning systems in which the feature extraction process is also optimized towards the task goal \cite{Lopez22}. The keyword spotting (KWS) task, which deals with the recognition of a very limited vocabulary in speech signals, is, indeed, not alien to this trend \cite{Simon20,Lopez21b,Peter22}. In KWS, researchers seek to replace solid \emph{handcrafted} speech features ---e.g., log-Mel features and Mel-frequency cepstral coefficients (MFCCs)--- by learnable features that are able to yield better KWS performance.

In spite of recent attempts \cite{Simon20,Lopez21b,Peter22}, the above goal has not been achieved. In \cite{Simon20}, Mittermaier \emph{et al.} studied the integration of a trainable filterbank ---the so-called SincNet \cite{Ravanelli18}--- into a convolutional neural network (CNN)-based KWS pipeline. In that system, the cut-off frequencies of a sinc-convolution-based filterbank are trained jointly along with the CNN acoustic model. While a formal comparison between SincNet and handcrafted features is missing in \cite{Simon20}, this analysis has recently been done in \cite{Peter22}, where Peter \emph{et al.} demonstrate that MFCCs remain superior for KWS.

In previous work \cite{Lopez21b}, we found no statistically significant differences between using learned filterbanks and log-Mel features for KWS. From this observation, we conjectured that much of the spectral information is redundant when it comes to the recognition of a set of few keywords\footnote{Although in the context of reduced-precision speech features for KWS, this same hypothesis was independently made in \cite{Riviello19}.}. Inspired by this, in this paper, we prove ---for the first time, to the best of our knowledge--- that filterbank learning outperforms handcrafted speech features for KWS as long as the number of filterbank channels is drastically reduced. While this might lead to certain KWS performance degradation, it can also yield a significant energy consumption and inference time reduction\footnote{This is as a result of the number of multiplications in the acoustic model directly depending on the size of the input feature matrix.}, which is of utmost importance when deploying typical always-on KWS on low-resource devices.

In this work, we conduct experiments on a noisy version of the Google Speech Commands Dataset \cite{Warden18,Lopez21} by using a CNN-based KWS system integrating residual connections \cite{Tang18}. Experimental results prompt the following observations:
\begin{enumerate}
	\item Filterbank learning adapts to noise spectral characteristics to offer a higher degree of robustness to noise;
	\item The use of dropout \cite{Dropout} enhances robustness to noise and generalization capabilities of learned filterbanks;
	\item Switching from 40-channel log-Mel features to 8-channel learned features yields a relative KWS accuracy loss of only 3.5\% while involving a 6.3$\times$ energy consumption reduction;
	\item Switching from 8-channel log-Mel features to 5-channel learned features allows us for essentially maintaining KWS accuracy while leading to a 2$\times$ energy consumption reduction.
\end{enumerate}

The rest of the paper is organized as follows. Section \ref{sec:method} presents our filterbank learning methodology. The experimental setup is described in Section \ref{sec:setup}. Results are shown and discussed in Section \ref{sec:results}, and Section \ref{sec:conclusions} concludes this work.

\section{Filterbank Learning Methodology}
\label{sec:method}

\begin{figure*}
		\begin{picture}(100,100)
		\put(0,0){\includegraphics[width=\linewidth]{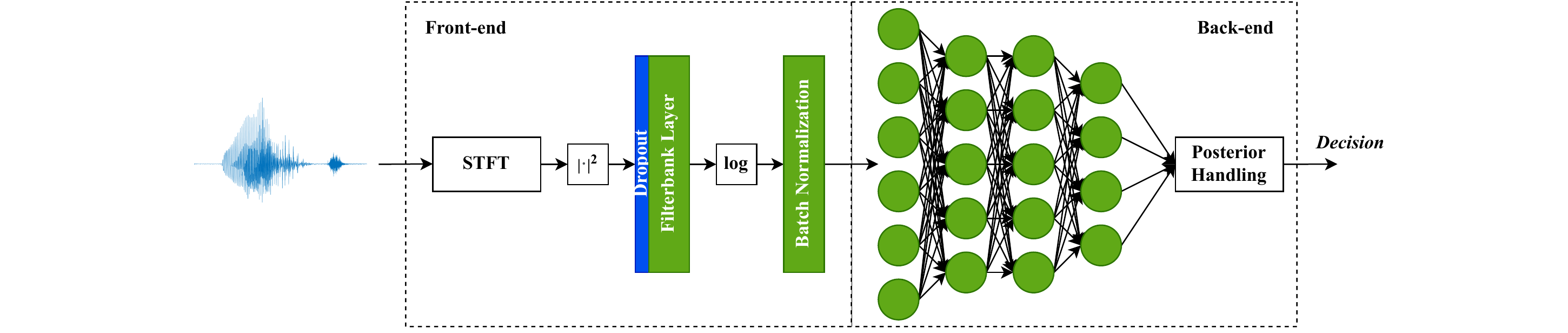}}
		\put(100,67){\small $x(m)$}
		\put(162,67){\small $X(t,f)$}
		\put(196,67){\small $\mathbf{X}$}
		\put(224,67){\small $\mathbf{Y}$}
		\end{picture}
	\caption{Block diagram of the deep keyword spotting system employed in this work. See the text for further details.}
	\label{fig:system}
\end{figure*}
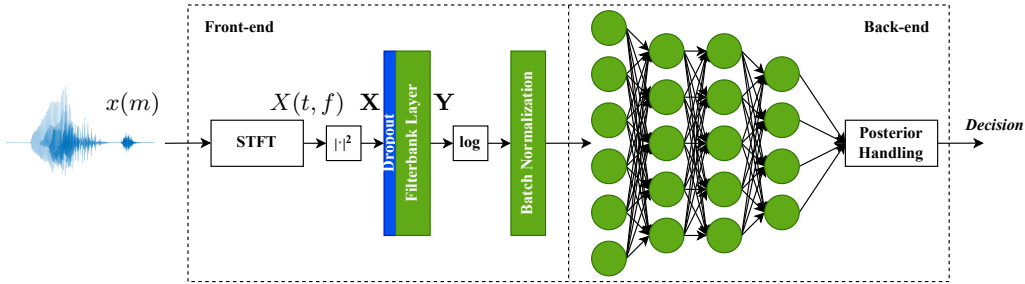

A block diagram of the KWS system employed in this work integrating filterbank learning can be seen in Fig. \ref{fig:system}. First, the short-time Fourier transform (STFT) $X(t,f)$ of a discrete-time input speech signal $x(m)$, potentially comprising a keyword, is calculated. Note that $t=1,2,...,T$ and $f=1,2,...,F$ indicate, respectively, the time frame and linear frequency bin. Furthermore, $T$ and $F$ denote the total number of time frames and linear frequency bins, respectively, in the input speech signal. We can define a $T\times F$ matrix $\mathbf{X}$ representing $x(m)$ in the linear power spectral domain as follows:
\begin{equation}
	\mathbf{X}=\left[\begin{array}{ccc}
		|X(1,1)|^2 & \cdots & |X(1,F)|^2 \\
		\vdots & \ddots & \vdots \\
		|X(T,1)|^2 & \cdots & |X(T,F)|^2
	\end{array}\right].
\end{equation}
Next, the filterbank layer ---optionally considering dropout \cite{Dropout} at training time--- implements the matrix multiplication 
\begin{equation}
	\mathbf{Y}=\mathbf{X}\cdot g(\mathbf{W}),
\end{equation}
where $\mathbf{W}\in\mathbb{R}^{F\times K}$ is the learnable $K$-channel filterbank matrix, which is optimized \emph{jointly} along with the acoustic model (i.e., back-end) by backpropagation. Moreover, $g(\cdot)=\max(\cdot,\;0)$ is the rectified linear unit function, which is element-wise applied to guarantee the positivity of the filterbank weights. The result of log-compressing the $T\times K$ filterbank representation $\mathbf{Y}$ as in $\log\left(\max\left(\mathbf{Y},\;\eta=e^{-50}\right)\right)$, where the operators $\log(\cdot)$ and $\max(\cdot)$ are element-wise applied, is input to a batch normalization layer producing the speech features for the acoustic model.

Notice that the only difference between this filterbank learning scheme and the one that we already proposed in \cite{Lopez21b} is the optional consideration of dropout \cite{Dropout} at training time, the goal of which is to improve robustness and generalization of individual filterbank channels.

For acoustic modeling, we use a deep residual CNN integrating dilated convolutions \cite{Tang18}, since in a recent overview article \cite{Lopez22} we concluded that state-of-the-art KWS acoustic modeling relies on CNNs incorporating both residual connections and a scheme to seize long time-frequency patterns. This model produces word-level posteriors from one-second long input speech segments $x(m)$. A keyword is spotted every time that its associated posterior probability is the largest among all posteriors.

\section{Experimental Setup}
\label{sec:setup}


\subsection{Noisy Google Speech Commands Dataset}
\label{ssec:ngscd}

For experimental purposes, we employ the noisy version of the Google Speech Commands Dataset (GSCD) \cite{Warden18} that was introduced in \cite{Lopez21}. This noisy GSCD defines training, validation and test sets composed of one-second long speech utterances ---comprising one word each--- that are contaminated by additive noises from the datasets NOISEX-92 \cite{Varga93} and CHiME-3 \cite{CHiME3b}. On the one hand, the types of noise present in the training and validation sets are vehicle interior, factory 1, bus, pedestrian street, white noise, babble, machine gun and F-16 cockpit. On the other hand, test set noise types are vehicle interior, factory 1, bus and pedestrian street (\emph{seen noises}), as well as factory 2, Buccaneer jet cockpit, caf\'e and street junction (\emph{unseen noises}). In addition, apart from the clean case also being considered, speech signals are contaminated at the following signal-to-noise ratios (SNRs): $\{0, 5, 10, 15, 20\}$ dB in the training and validation sets, and $\{-10, -5, 0, 5, 10, 15, 20\}$ dB in the test set. The training, validation and test sets contain 3,699, 427 and 497 utterances, respectively, from each combination of noise type and SNR level. The reader is referred to \cite{Lopez21} for further details about this noisy version of the GSCD.

\subsection{Configuration and System Training}
\label{ssec:training}

\subsubsection{Front-End}
\label{sssec:frontend}

For STFT computation, an analysis window of 30-ms duration ---corresponding to $N=480$ samples at a 16 kHz sampling rate--- with a 10-ms skip is employed. Hence, $F=(N/2)+1=241$ is the total number of linear frequency bins. In addition, when dropout is considered, the dropout rate is 0.4. Departing from the standard number of filterbank channels in the KWS literature $K=40$ \cite{Lopez21,Lopez21b,Kim21,Lopez22}, different KWS systems with a reduced number of filterbank channels are trained to assess prospective benefits of filterbank learning. Note that $\mathbf{W}$ is initialized by a Mel filterbank.

\subsubsection{Back-End}
\label{sssec:backend}

The CNN back-end is trained to model 11 different classes: the 10 standard keywords of the GSCD ``yes'', ``no'', ``up'', ``down'', ``left'', ``right'', ``on'', ``off'', ``stop'' and ``go'' \cite{Warden18}, plus the filler (i.e., non-keyword) class. As is typical in the context of GSCD, all classes are close to be balanced across the training, validation and test sets \cite{Warden18,Lopez22}.

The CNN acoustic model is trained by means of Adam \cite{Adam} towards minimizing cross-entropy loss. The size of the mini-batch is 64 samples and early-stopping (monitoring the validation loss) with a patience of 5 epochs is used for regularization purposes. Finally, note that every KWS experiment is repeated 5 times by training 5 different acoustic models with different random initialization of their parameters in order to make sound conclusions from statistical tests.

\section{Results and Discussion}
\label{sec:results}

\begin{table*}[th]
	\caption{Average keyword spotting accuracy results (\%), from using log-Mel and learned features, as a function of the number of filterbank channels $K$. Accuracy values are broken down by SNR as well as by seen and unseen noises during the training phase. The number of multiplications of the acoustic model, which depends on the number of filterbank channels, is also shown. Statistically significant improvements with respect to log-Mel (Learned) are indicated in boldface (underlined).}
	\label{tab:results}
	\centering
	\setlength{\extrarowheight}{3pt}
	\resizebox{\linewidth}{!}{\begin{tabular}{ccccccccccc|cccccccc}
			\toprule
			\multicolumn{1}{c}{\textbf{\#Ch.}} & \textbf{\#Mult.} & \textbf{Method} & \multicolumn{8}{c}{\textbf{SNR (dB) - Seen Noises}} & \multicolumn{8}{c}{\textbf{SNR (dB) - Unseen Noises}} \\ \cline{4-11} \cline{12-19}
			\multicolumn{1}{c}{$K$} & & & \emph{-10} & \emph{-5} & \emph{0} & \emph{5} & \emph{10} & \emph{15} & \emph{20} & \emph{Clean} & \emph{-10} & \emph{-5} & \emph{0} & \emph{5} & \emph{10} & \emph{15} & \emph{20} & \emph{Clean} \\ \midrule
			\multicolumn{1}{c}{} & & log-Mel & 51.26 & 67.61 & 82.27 & 90.48 & 93.43 & 94.08 & 95.36 & 96.09 & 37.17 & 61.38 & 78.86 & 86.87 & 91.92 & 93.66 & 94.58 & 96.13 \\
			\multicolumn{1}{c}{40} & 895M & Learned & 49.61 & 66.93 & 82.46 & 90.58 & 93.96 & 94.23 & 95.41 & 95.58 & 35.41 & 58.84 & 77.94 & 86.82 & 92.24 & 93.47 & 94.70 & 95.60 \\
			\multicolumn{1}{c}{} & & Learned+D & 51.21 & 68.16 & 83.50 & 90.29 & 93.89 & 93.84 & 95.43 & 95.85 & 38.14 & \underline{63.29} & 78.96 & 87.35 & 92.29 & 93.86 & 94.90 & 96.20 \\ 
			\midrule
			\multicolumn{1}{c}{} & & log-Mel & 45.41 & 63.09 & 79.30 & 87.85 & 91.93 & 92.90 & 94.59 & 95.70 & 32.00 & 55.67 & 74.34 & 83.94 & 90.23 & 92.29 & 93.91 & 95.07 \\
			\multicolumn{1}{c}{10} & 188M & Learned & 47.37 & 64.15 & 80.10 & 88.86 & 92.71 & 93.55 & 94.88 & 95.87 & 33.54 & 55.91 & 74.85 & 84.67 & 91.22 & 92.48 & 94.22 & 95.31 \\
			\multicolumn{1}{c}{} & & Learned+D & 46.72 & 63.57 & 80.39 & 88.36 & 92.13 & 93.29 & 94.71 & 95.58 & 32.50 & 57.39 & 75.14 & 84.57 & 90.67 & \textbf{93.23} & 94.39 & 95.72 \\
			\midrule
			\multicolumn{1}{c}{} & & log-Mel & 44.64 & 60.51 & 76.01 & 85.00 & 90.29 & 91.88 & 93.31 & 94.59 & 28.25 & 49.94 & 68.92 & 80.82 & 87.45 & 90.40 & 92.12 & 93.93 \\
			\multicolumn{1}{c}{8} & 141M & Learned & 45.70 & 62.22 & 77.97 & 86.06 & 90.75 & 91.45 & 92.97 & 94.15 & 29.65 & \textbf{53.11} & \textbf{72.24} & \textbf{84.04} & \textbf{90.33} & \textbf{92.16} & \textbf{93.40} & 94.24 \\
			\multicolumn{1}{c}{} & & Learned+D & 46.23 & \textbf{63.67} & 78.99 & \underline{\textbf{87.58}} & 90.94 & 92.32 & 94.13 & \underline{\textbf{95.70}} & 30.52 & \textbf{55.02} & \textbf{73.76} & \textbf{84.62} & \textbf{90.45} & \underline{\textbf{92.94}} & \textbf{94.00} & \textbf{94.82} \\
			\midrule
			\multicolumn{1}{c}{} & & log-Mel & 42.90 & 60.72 & 76.28 & 84.49 & 89.18 & 90.31 & 92.51 & 94.23 & 27.88 & 48.20 & 68.51 & 80.53 & 87.06 & 90.40 & 92.72 & \underline{94.82} \\
			\multicolumn{1}{c}{7} & 118M & Learned & 44.08 & 62.27 & 76.91 & 85.53 & 89.54 & 91.01 & 92.68 & 94.49 & 30.47 & \textbf{51.80} & \textbf{72.12} & 82.08 & 88.05 & 90.79 & 92.53 & 93.74 \\
			\multicolumn{1}{c}{} & & Learned+D & 45.53 & 61.35 & 78.00 & \underline{\textbf{87.00}} & \underline{\textbf{90.80}} & \textbf{91.43} & 93.26 & 94.88 & 31.32 & \underline{\textbf{54.68}} & \underline{\textbf{73.30}} & \underline{\textbf{83.94}} & \underline{\textbf{89.50}} & \underline{\textbf{92.09}} & 92.99 & 94.32 \\
			\midrule
			\multicolumn{1}{c}{} & & log-Mel & 40.12 & 56.52 & 73.31 & 82.56 & 87.13 & 88.41 & 89.78 & 92.13 & 24.52 & 46.84 & 66.26 & 79.93 & 85.03 & 88.59 & 90.59 & 92.50 \\
			\multicolumn{1}{c}{5} & 71M & Learned & \textbf{42.92} & \textbf{59.40} & 75.34 & \textbf{84.69} & 88.26 & 89.57 & \textbf{91.52} & \textbf{93.55} & \textbf{28.34} & \textbf{50.47} & \textbf{70.11} & 80.82 & \textbf{86.99} & 89.41 & 91.22 & 92.74 \\
			\multicolumn{1}{c}{} & & Learned+D & \underline{\textbf{44.66}} & \underline{\textbf{61.76}} & \underline{\textbf{76.93}} & \textbf{84.57} & \textbf{88.77} & \textbf{90.07} & \textbf{91.79} & \textbf{93.91} & \textbf{28.39} & \textbf{51.29} & \textbf{70.04} & \textbf{82.54} & \textbf{87.67} & \underline{\textbf{91.27}} & \textbf{92.26} & \textbf{93.81} \\
			\bottomrule
	\end{tabular}}
\end{table*}

In this section, we compare the use of log-Mel features with the utilization of the learnable features described in Section \ref{sec:method} with ---Learned+D--- and without ---Learned--- dropout. Table \ref{tab:results} shows KWS accuracy results (averaged across 5 experiment repetitions), in percentages, as a function of the number of filterbank channels $K$. \emph{Due to space constraints}, we only report results for a more relevant selection of $K$ values (i.e., $\{40, 10, 8, 7, 5\}$). Statistically significant accuracy improvements are identified by means of two-sample $t$-tests \cite{ttest} with a significance level of 0.05. In Table \ref{tab:results}, statistically significant improvements with respect to log-Mel (Learned) are indicated in boldface (underlined).

Table \ref{tab:results} also shows the number of multiplications of the acoustic model per second of input speech, which highly depends on $K$ due to the use of residual connections. Because \emph{the number of multiplications exhibits a strong positive linear relationship with the energy consumption} of the KWS system, ($R^2=0.9641, p=0.0001$) \cite{Raphael18}, we use the former as a proxy for the latter in this work.

Consistent with what we determined in \cite{Lopez21b}, from Table \ref{tab:results}, we see no statistically significant performance differences between log-Mel and learned features when employing a standard number of filterbank channels (i.e., $K=40$). However, filterbank learning starts to become beneficial when $K<10$, especially when dealing with unseen noise types. In particular, for $K=5$, Learned+D outperforms log-Mel features in all the evaluated noisy conditions. We conjecture that the combination of dropout and a reduced $K$ creates a bottleneck in the filterbank layer (quite similar to the ones observed in autoencoders), thereby providing a better knowledge representation for filterbank learning.

\begin{figure}
	\begin{center}
		\includegraphics[width=\linewidth]{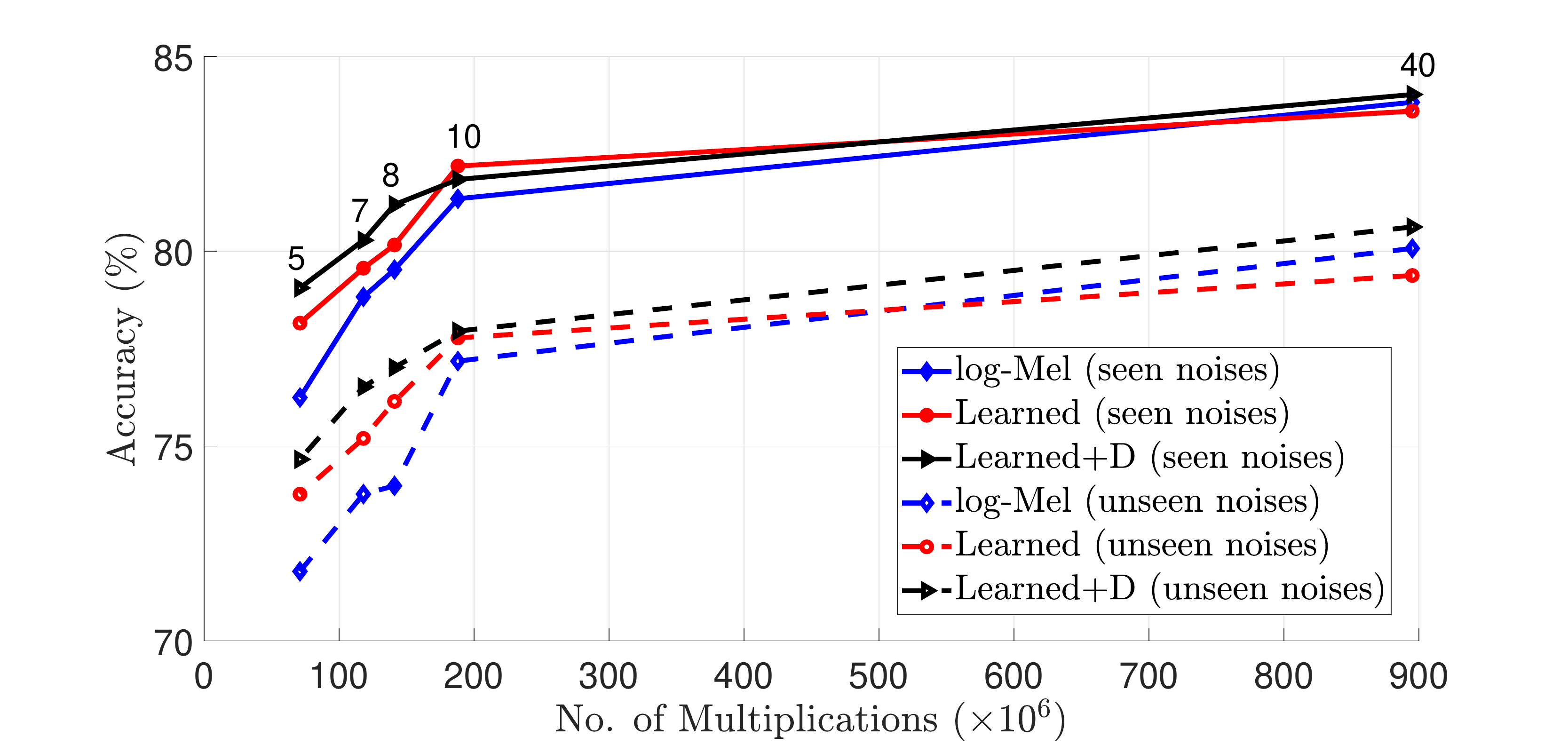}
	\end{center}
	\caption{Keyword spotting accuracy (\%) averaged across SNRs, from using log-Mel and learned features, as a function of the number of multiplications of the acoustic model. The corresponding number of filterbank channels, $K$, is indicated above the curves.}
	\label{fig:results}
\end{figure}

While it is true that KWS performance tends to drop when reducing $K$, filterbank learning cushions this drop while achieving noticeable energy savings. For example, switching from standard 40-channel log-Mel (avg. accuracy of 81.95\%) to Learned+D when $K=8$ (avg. accuracy of 79.11\%) leads to a relative KWS accuracy loss of only 3.5\% while involving a $895\mbox{M}/141\mbox{M}\approx 6.3\times$ energy consumption reduction\footnote{For comparison, the relative KWS accuracy loss between 40-channel and 8-channel log-Mel (avg. accuracy of 76.75\%) is of 6.3\%.}. Furthermore, switching from 8-channel log-Mel (avg. accuracy of 76.75\%) to Learned+D when $K=5$ (avg. accuracy of 76.86\%) allows us to essentially maintain KWS performance while yielding a reduction in energy consumption of a factor $141\mbox{M}/71\mbox{M}\approx 2$. To assess at a glance the more robust behavior of filterbank learning to the decreasing of $K$ (for both seen and unseen noises), Fig. \ref{fig:results} plots KWS accuracy (averaged across SNRs) as a function of the number of multiplications of the acoustic model.

\subsection{Filterbank Examination}
\label{ssec:analysis}

\begin{figure}
	\begin{subfigure}{\linewidth}
		\centering
		\includegraphics[width=\linewidth]{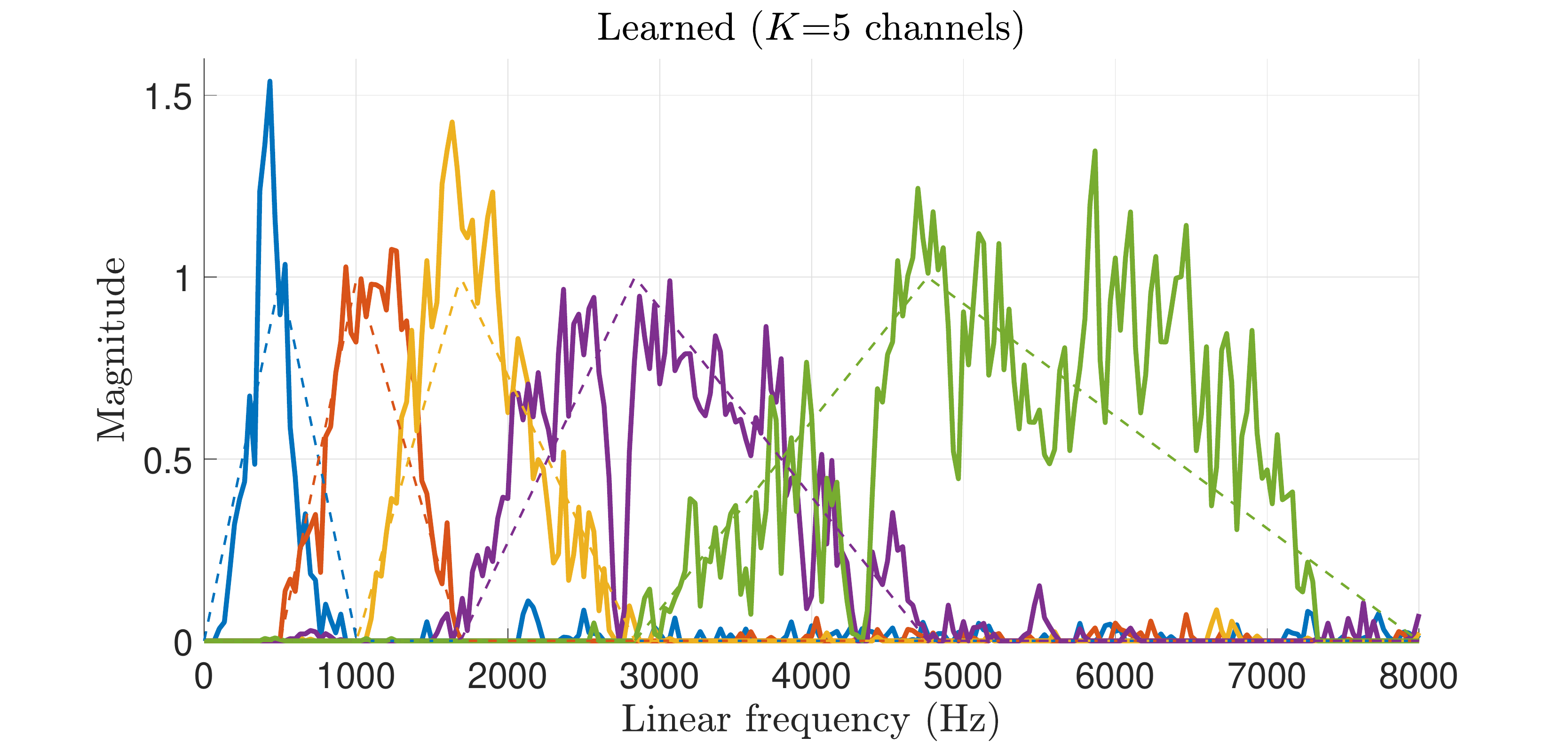}
		\caption{}
		\label{fig:a}
	\end{subfigure}\\%
	\begin{subfigure}{\linewidth}
		\centering
		\includegraphics[width=\linewidth]{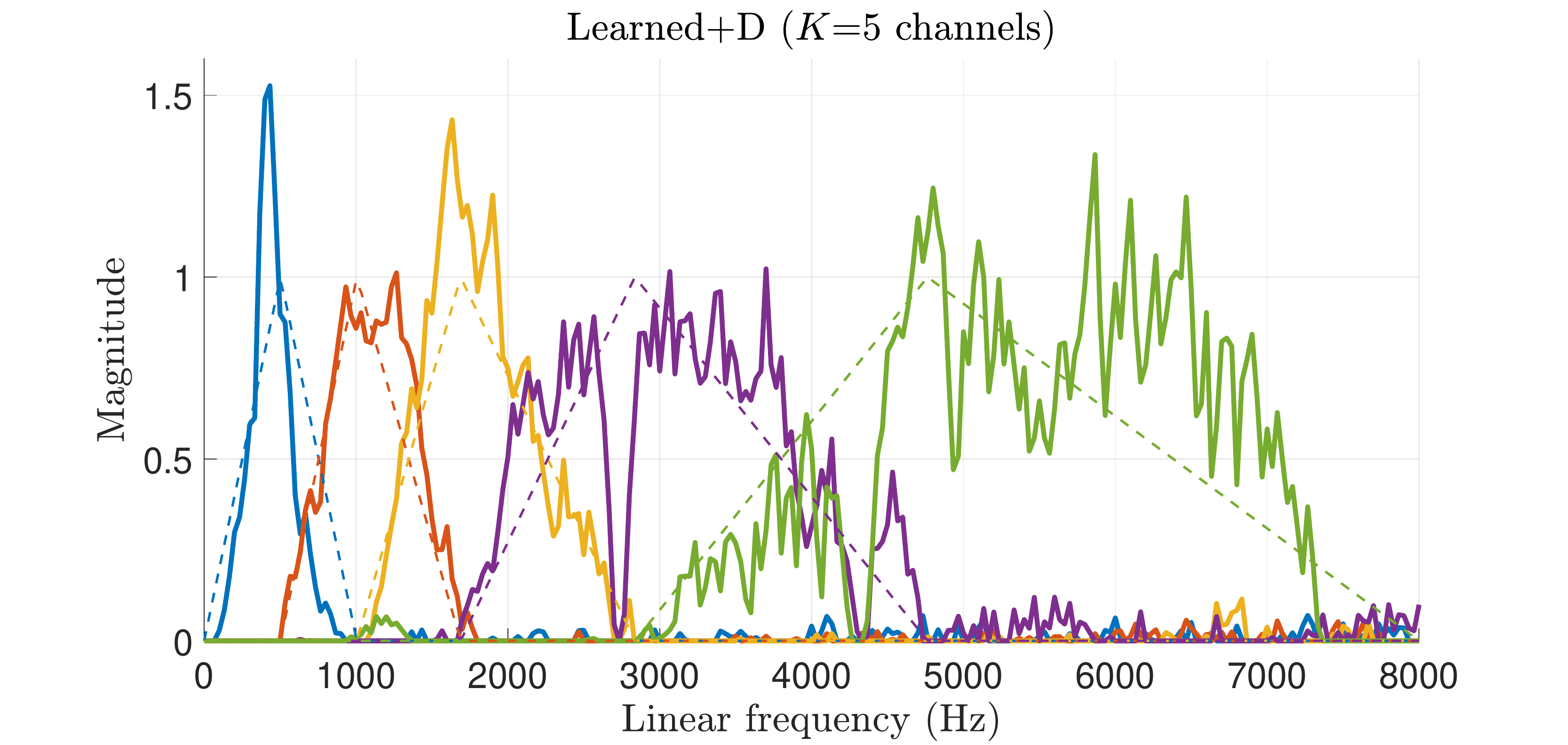}
		\caption{}
		\label{fig:b}
	\end{subfigure}
	\caption{Average (across the 5 experiment repetitions) learned filterbanks (solid lines) when considering $K=5$ channels: (a) without dropout and (b) with dropout. For the sake of comparison, 5-channel Mel filterbanks (dashed lines) are overlaid.}
	\label{fig:learned_fb}
\end{figure}

Figure \ref{fig:learned_fb} depicts the average (across the 5 experiment repetitions) 5-channel learned filterbanks with (bottom) and without (top) dropout. As can be seen from this figure, the inclusion of dropout produces a smoother version of the filterbank learned without dropout, especially at lower frequencies.

It is interesting to note how learned filterbanks suppress frequency components near 2.7 kHz and 4.3 kHz. We pinpoint that this is due to the long-term spectrum of the F-16 cockpit training noise exhibiting strong peaks at those frequencies. In addition, the learned filterbanks seem to give more emphasis to the frequency range 5.8-7 kHz than the Mel filterbank (dashed lines in Fig. \ref{fig:learned_fb}). This may be due to the fact that most of the training noise long-term spectra have fading tails at higher frequencies, so speech information potentially useful for KWS is less distorted in the 5.8-7 kHz range. In conclusion, filterbank learning seems to adapt to noise spectral characteristics to offer a higher degree of robustness to both seen and unseen noises.

It is remarkable that this analysis, illustrated by considering $K=5$, holds also valid for $K>5$, with similar plots to those in Fig. \ref{fig:learned_fb}. Therefore, we believe that the potential advantages of filterbank learning adapting to noise characteristics are masked by the aforementioned information redundancy issue \cite{Riviello19,Lopez21b} when $K\ge 10$ (see Table \ref{tab:results}).

\section{Concluding Remarks}
\label{sec:conclusions}

For the first time, to the best of our knowledge, we have shown in this study that, when the number of filterbank channels is substantially decreased, filterbank learning is able to outperform handcrafted speech features for KWS. The entailed KWS performance-energy consumption trade-off is further compensated by using dropout for filterbank learning, which might be of particular interest to practitioners who deploy common always-on KWS on low-resource devices.

Future work envisages a comprehensive study on this matter aiming at obtaining deeper insight helping to design superior speech features for robust, small-footprint KWS.

\bibliographystyle{IEEEbib}
\bibliography{strings,refs}

\end{document}